\documentclass{article}

\usepackage{graphicx}

\begin{document}

\hbox{}
\nopagebreak
\vspace{-3cm}
\begin{flushright}
{\sc OUTP-99 52P} \\
{\sc CPT-P.3891} \\

\end{flushright}

\vspace{1in}

\begin{center}
{\Large \bf Spatial 't Hooft loop, hot QCD and $Z_N$ domain walls}
\vspace{0.1in}

\vspace{0.5in}
{\large C. Korthals-Altes$^1$,  A. Kovner$^{1,2}$ and M. Stephanov$^3$}\\
\vspace{.4in}
{\small
$^1$Centre Physique Theorique au CNRS, Case 907, Luminy\\
F13288, Marseille Cedex, France\\
$^2$Theoretical Physics, University of Oxford, 1 Keble Road, \\
Oxford, OX1 3NP, UK}\\
$^3$Department of Physics, University of Illinois,\\
845 West Taylor Street, Chicago, IL 60607, USA\\
\vspace{0.5in}

{\bfseries\sc  Abstract}\\
\end{center}
\vspace{.2in}
We show that the deconfinement phase transition in the pure Yang-Mills theory can be 
characterized by the change of behaviour the spatial 't Hooft loop, $V(C)$. 
In the confining phase $V$ has a perimeter law behaviour $V(C)\propto \exp\{-mP(C)\}$,
while in the deconfined phase it has the area law behaviour
$V(C)\propto \exp\{-\alpha S(C)\}$. We show that the area law behaviour of the 
't Hooft loop is intimately related to the 
plasma-like distribution of the color charges in the hot QCD vacuum. We also show 
that the ``dual string tension'' $\alpha$ is equal to the``wall tension''of the $Z_N$
domain walls 
previously calculated in \cite{wall}.
All these properties generalize immediately to other nonabelian theories without 
fundamental charges, such as supersymmetric Yang Mills. In theories with 
fundamental charges the 't Hooft loop presumably has 
the area law behaviour already at zero
temperature and therefore is not a good order parameter in the strict sense.

\vfill

\newpage

The deconfining phase transition in QCD is a subject with long history. It has seen
a lot of activity in recent years primarily motivated by its possible relation
with the heavy ion collision experiments at RHIC. 
Although the direct phenomenological relevance of the equilibrium high temperature QCD
at RHIC is only conjectural, the subject of finite temperature behaviour of nonabelian
theories is very fascinating theoretically and has possible cosmological applications
in particular in relation to baryogenesis.

In this paper we will concentrate on a certain theoretical aspect of the hot phase
in nonabelian gauge theories without fundamental matter fields.
This class of theories is believed to have a deconfining phase transition to a 
``color plasma phase''. Starting with early discussions of universality 
\cite{larry},\cite{ben} it has been studied quite 
extensively by both analytical perturbative \cite{perturb} 
and lattice nonperturbative methods \cite{lattice}
and the nature of the phase transition is fairly well established. 

Despite a lot of work on the subject which yielded very interesting results, it is 
perhaps surprising that
a description of the phase transition in terms of a canonical order parameter does not exist.
Although one frequently refers to the Polyakov line $P={\rm Tr} P\exp\{ig\int d t A_0\}$
as an order parameter, it does not have the same status as the magnetisation 
in the Ising or Heisenberg model.
While magnetisation is a canonical operator in the Hilbert space of the theory, 
the Polyakov line is not. It appears as an auxiliary quantity in the path 
integral formulation of the statistical sum, and while indeed it has 
qualitatively different behaviour in the confined and the deconfined phases, 
the ``nonvanishing expectation value of $P$'' above the phase transition
can not be directly related
to nonvanishing expectation value of any physical operator.

This ambiguity in the status of $P$  several years ago has prompted discussions on 
the physical status of the ``domain walls'' - field configurations which interpolate
between different ``vacuum'' values of $P$ in the path integral 
\cite{wallsfor},\cite{wallsagainst}.
The outcome of this discussion has been itself 
somewhat unsatisfactory. On one hand as a result it is now
understood that the ``domain walls'' are not physical objects of the same type as
domain walls in the Ising model. They are not 
field configurations which interpolate
between different physical states at spatial infinity. On the other hand 
the physical meaning of these walls has not been clarified. In particular
it would be interesting to relate the ``wall tension'' (which 
has been calculated perturbatively \cite{wall} and 
nonperturbatively \cite{lwall},\cite{mike})
to the expectation value of some physical observable.

The aim of this note is to rectify this situation. We will show that there indeed is
a well defined order parameter whose behaviour changes qualitatively across the
phase transition. This order parameter is the expectation value of the disorder 
variable - the spatial 't Hooft loop \cite{thooft}. The operator $V(C)$ creates an 
elementary 
flux of magnetic field along a closed contour $C$. In the low temperature phase
the VEV of $V$ for large contours 
falls off according to a perimeter law, while in the high temperature phase 
it has an area law behaviour 
\begin{equation}
<V(C)>\propto \exp\{-\alpha S(C)\}
\end{equation}
The ``dual string tension'' $\alpha$ turns out to be precisely given by the 
``wall tension'' of the $Z_N$ domain wall. 
In this way the wall tension is directly related to a physical observable\footnote{We
note that the 2+1 dimensional analog of the 't Hooft loop was discussed in the 
context of the finite temperature gauge theory in \cite{mike}. However the
change of its behaviour across the phase transition was not analysed in that work.}.
We hasten to add that the fact that the spatial 'tHooft 
loop has an area law does not mean that magnetic charges in the theory 
are confined by a linear potential. Just like the area law of 
the spatial Wilson loop does not indicate confinement of electric charges.

Let us start with recalling the definition of the disorder variable 
$V$ \cite{thooft}. 
In the following we will for definiteness consider the $SU(2)$ gauge group and
will discuss the generalization to other groups later.
The defining property of $V$ is that it satisfies the following commutation relation with
the spatial fundamental Wilson loop $W$
\begin{equation}
V^\dagger(C)W(C')V(C)=e^{i\pi n(C,C')}W(C')
\end{equation}
where $n(C,C')$ is the linking number of the curves $C$ and $C'$.

The operatorial representation of $V$ is constructed as follows.
Consider a closed contour $C$ which lies in the $xy$ plane. Define the function
$\omega_C(x)$ which is equal to the solid angle subtended by $C$ as seen from the 
point $x$. The function $\omega$ is continuos everywhere, except on a surface $S$
bounded by $C$, where it jumps by $4\pi$. Other than the fact that $S$ is bounded by 
$C$, its location is arbitrary. Now consider 
the operator of the ``singular gauge transformation'' in the third
colour direction with the gauge function $\omega/2g$ 
\begin{equation}
V(C)=\exp\{{i\over 2g}\int d^3x (\bar D^i_{3a})\omega_C E^i_a\}
\label{v1}
\end{equation}
The bar over the covariant derivative indicates the fact that 
only the regular part of the derivative of $\omega$ enters in the definition of the
operator $V$. An alternative, and 
a somewhat simpler form of $V$ is obtained by using the fact that on the physical 
states $D^iE^i=0$ and that the solid angle $\omega_C(x)$ vanishes at infinity.
Therefore on physical states
\begin{equation}
V(C)=\exp\{{2\pi i\over g}\int_S d^2 S^i E^i_3\}
\label{v}
\end{equation}
The integration here is over the surface $S$ on which $\omega$ is defined to 
have the discontinuity.
If no fields in the fundamental representation are present in the theory, 
the operator $V$ does not depend on this 
surface, but rather depends only on its boundary $C$. To see this, note that 
changing $S$ to $S'$ adds to the phase ${2\pi\over g}\oint_{S-S'}d^2 S^iE^i$. In a
theory with only adjoint charges the charge within any closed volume is a
multiple integer of the gauge coupling $ \oint_{S-S'}d^2 S^iE^i=gn$
and therefore the extra phase factor is always unity.
In the following we will for simplicity always choose the surface $S$ to lie 
in the $xy$ plane.

It is easy to see that
\begin{equation}
V^\dagger(C)A^a_i({\bf x})V(C)=A^a_i({\bf x})+a^a_i({\bf x})
\end{equation}
with
\begin{equation}
a^a_i({\bf x})=
\delta^{a3}\delta_{i3}{2\pi\over g}\delta({\bf x}-S)
\label{a}
\end{equation}
The $\delta$-function in this equation 
is one dimensional and is defined such that its integral along a curve normal to 
the surface $S$ is equal to unity. 
 
The operator $V(C)$ is therefore seen to create an infinitely thin elementary 
vortex of magnetic
field in the third color direction along the curve $C$.
Clearly our choice of the third direction in the color space is arbitrary. One
can equaly well consider any other direction. Different operators defined in this 
way transform into each other under the gauge transformations and therefore are 
identical when acting on physical states.

The operator $V$ was introduced by 't Hooft in order to study the phase structure of
gauge theories. 
In the zero temperature ground state of the pure Yang-Mills theory the Wilson loop
has an area law behaviour.
't Hooft's general analysis showed that (unless the gauge symmetry 
was partially broken) the area law behaviour of 't Hooft loop and Wilson loop were
mutually exclusive. 
The 't Hooft loop therefore has a perimeter law behaviour in the ground state.
This statement however pertains only to the zero temperature ground
state, which is Lorentz invariant and where the spacelike and timelike loops behave in the same way. 
Our purpose here is to study the expectation value of the 
't Hooft loop at high temperature.

The expectation value of the 't Hooft loop at finite temperature is given by the 
following expression
\begin{equation}
<V(C)>={\rm Tr} e^{-{\beta\over 2}(E^2+B^2) }e^{i{2\pi i\over g}\int d^2 S^i E^i_3}
\label{vt}
\end{equation}
Note that formally
eq.(\ref{vt}) is the same as for the partition function except that the Hamiltonian
has an extra term linear in the electric field
\begin{equation}
\delta H=-iT\int d^3x a^i_aE^i_a
\end{equation}
with $a^i_a$ defined in eq.(\ref{a}) and $T={1\over\beta}$.
With the help of the standard manipulations it is easy to cast this expression in the
path integral form. Introducing the imaginary time axis and the 
Lagrange multiplier field $A_0$ in the standard way
we obtain
\begin{equation}
<V>=\int DA_iDA_0\exp\{-{1\over 2}\int_0^{\beta} dt\int d^3x
\left(\partial_0A_i^a-(D_iA_0)^a-Ta_i^a\right)^2+(B^a)^2\}
\end{equation}

Our task is now simple. In this expression the ``external field'' $a_i$ enters 
only in one place - it shifts the spatial derivative 
of the zero Matsubara frequency 
mode of $A_0$.
The integration of the nonzero Matsubara modes therefore proceeds in precisely the 
same way as in the standard calculation 
of the finite temperature effective potential \cite{weiss},\cite{korthals}.
In fact we can take the whole page from the book of \cite{korthals}
and derive the constrained
effective Lagrangian for the zero Matsubara frequency modes in the presence of 
the external field $a_i$. It is obvious that the calculation is identical to that of
\cite{korthals}. For simplicity we will restrict ourselves to the one loop result
\begin{equation}
S_{eff}={2T^2\over g^2}(\partial_iq+{g\over 2}
a_i)^2+{4\over 3}T^4q^2(1-{q\over\pi})^2
\label{action}
\end{equation}
Here $q$ is defined (\cite{korthals}) as the average value 
of the first eigenvalue 
of the matrix $A_0={A^a_0\tau^a\over gT}$ at zero Matsubara frequency.
The matrices 
$\tau^a$ are the generators of $SU(2)$ in the fundamental representation
and are normalized according to ${\rm tr}\tau^a\tau^b={1\over 2}\delta^{ab}$.

To find the average of $V$ we have to find the configuration of $q$ 
which minimizes the action eq.(\ref{action}).
Qualitatively the form of the solution is clear.
Let us consider  a large 't Hooft loop
such that its radius is much larger 
than the electric mass in eq.(\ref{action}). Clearly very far from the loop at
spatial infinity
the field $q$ must take the value which minimizes the potential term in 
eq.(\ref{action}). There are two such values 
$q=0,\pi$. For definiteness we take the asymptotic value at infinity to be $0$.
On the other hand on the surface $S$, where the external field
$a_i$ is a delta function, $q$ has to jump by $\pi$. It is also clear that 
for a large loop the solution must be practically $x$ and $y$ independent
as long as $x$ and $y$ are well within the surface $S$.
For these values of $x$ and $y$ therefore we are looking for
the $z$ - dependent solution which everywhere except at $z=0$ is a solution of free
equations (the source vanishes), while at $z=0$ jumps by $\pi$
and is constrained to approach $0$ at $z\rightarrow\pm\infty$.
But of course we know what this 
solution is. Recall that the action eq.(\ref{action}) without the source term 
allows  wall-like solutions $q_{wall}$. They  interpolate 
between $q=0$ at $z\rightarrow -\infty$ and $q=\pi$ at $z\rightarrow+\infty$. 
This is precisely the $Z_2$ wall alluded to earlier. Let us pick
the profile which corresponds to the wall at $z=0$. This means that
$q_{wall}(0)={\pi\over 2}$. We can now construct the
following function
\begin{eqnarray}
q_S(z)=&q_{wall}(z)&,\ \ \ \ z<0\\
&q_{wall}(z)&-\pi,\ \ \ \ z>0 \nonumber
\label{solution}
\end{eqnarray}
This function has precisely the required properties: it satisfies the correct
boundary conditions, has the correct discontinuity 
and solves the sourceless equations everywhere except at discontinuity.
Close to the boundary of $S$ the solution will be of course modified - it is not
$x,y$ independent anymore. We will not address here the question of what the
exact form of the solution is\footnote{We only note that disregarding the boundary
would give one infinity of wall solutions shifted with 
respect to each other in the $z$ direction. The main effect of the boundary terms is to 
fix the wall centered on the plain containing the contour $C$.}. 
It is however clear that the behaviour of $q$ close to
the boundary of $S$ does not affect the leading contribution to the action. This
leading contribution for the large 't Hooft loop is obviously proportional to
the minimal area $S_m$ subtended by the contour $C$

\begin{figure}
\includegraphics[angle=270]{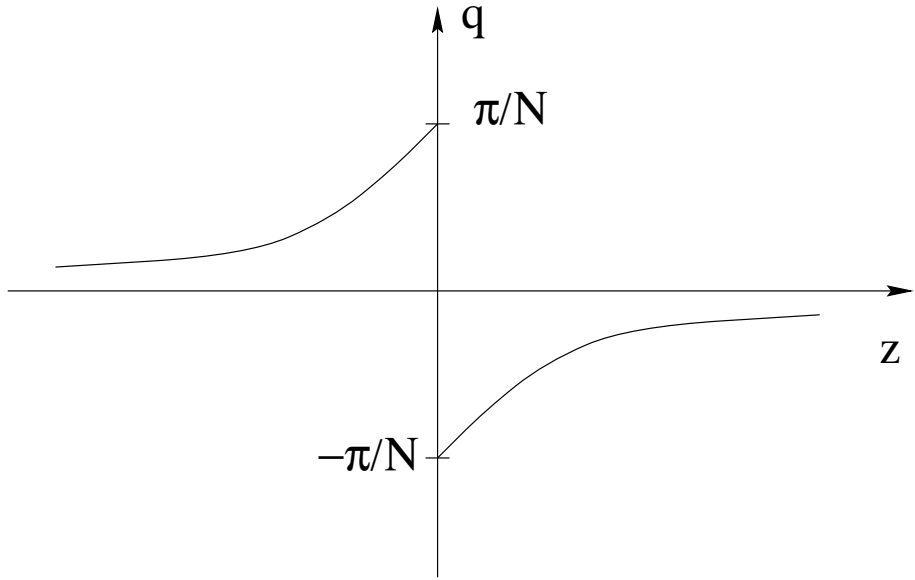}
\label{fig.1}
\caption{The profile of the field $A_0$ that dominates the steepest descent calculaltion
of $V$.}
\end{figure}
\begin{equation}
S_{eff}=\alpha S_m
\end{equation}
where $\alpha$ is precisely the wall tension of the $Z_2$ wall.

Note that the action is proportional to the minimal area $S_m$ subtended by the
contour $C$ rather than the area of the 
surface $S$ which appears explicitly in the definition of the operator $V$ and eq.(\ref{a}). 
Independence on $S$ is of course a consequence of the fact mentioned earlier, that 
the operator $V$ is independent of $S$ in the absence of fundamental charges.
In the framework of our path integral calculation this is easy to understand. 
Let us consider the operator $V$ defined on some other surface defined by 
equation $z=f(x,y)$ rather than $z=0$, but which has the
same boundary\footnote{For simplicity, here and in the rest of this note we consider
only surfaces which are smooth on the scale of the electric screening length.}. 
This means that the source term in eq.(\ref{action}) is modified accordingly and 
induces the jump by $\pi$ on
this new surface. The solution to the equations of motion with the required singularity structure
now is
\begin{eqnarray}
q_S(z)=&q_{wall}(z)&,\ \ \ \ \ \ \ z<f(x,y)\\
&q_{wall}(z)&-\pi,\ \ \ \ z>f(x,y) \nonumber
\label{solution1}
\end{eqnarray}
Since the potential term in eq.(\ref{action}) is symmetric under shift of $q$ by $\pi$,
the action of this solution is 
obviously the same as of eq.(\ref{solution}) and is proportional
to the minimal area $S_m$.

We therefore conclude that the expectation value of a large 't Hooft loop follows an
area law
\begin{equation}
<V(C)>=e^{-S_{eff}}=e^{-\alpha S}
\end{equation}

The generalization of the preceding discussion to $SU(N)$ gauge theories is straightforward. The
object of interest here is the 't Hooft loop
\begin{equation}
V(C)=\exp\{{ i\over gN} \int d^3x Tr(\bar D^i\Omega_C) E^i\}
\label{v2}
\end{equation}
with $\Omega_C(x)$, as before, the singular (solid angle) gauge function 
in the hypercharge direction
\begin{equation}
\Omega_C(x)=\omega_C(x)Y
\end{equation}
where the hypercharge generator $Y$ is defined as
\begin{equation}
Y={\rm diag} \left(1,1,...,-(N-1)\right)
\end{equation}

The calculation of $<V(C)>$ proceeds exactly along the same lines as before. 
In the path integral representation\footnote{We have switched to the standard
matrix notation where $E=E^a\lambda^a$ with the fundamental representation
$SU(N)$ generators normalized according to $Tr\lambda^a\lambda^b={1\over 2}\delta^{ab}$.}
\begin{equation}
<V>=\int DA_iDA_0\exp\left\{-\int_0^{\beta} dt\int d^3x
Tr\left[\left(\partial_0 \vec A-\vec D A_0-T\vec a\right)^2+
(\vec B)^2\right]\right\}
\end{equation}
with
\begin{equation}
 a^i({\bf x})=
{2\pi\over {gN}}Y\delta^{i3} \delta({\bf x}-S)
\end{equation}

The effective action is easily calculated again following \cite{korthals}. The 
result is simple for the configurations of the vector potential of the form
\begin{equation}
A_0={qT\over g}Y
\end{equation}
The one loop result is
\begin{equation}
S_{eff}={T^2\over g^2}N(N-1)\left(\partial_iq(x)-{2\pi\over N}\delta^{i3}\delta(x-S)\right)^2+
{N^2(N-1)\over 3}T^4q^2\left(1-{Nq\over 2\pi}\right)^2
\end{equation}
One again obtains the area
law for $<V(C)>$  with the dual 
string tension given by the wall tension of the $Z_N$ ``domain wall''.

The following comment is in order here. As we mentioned earlier 
't Hooft gave an argument to the effect that in a
Lorentz invariant state the Wilson loop and the 't Hooft loop can not have simultaneously
an area law behaviour. As is well 
established, the spatial Wilson loop
at high temperature in fact does have an area law. Nevertheless  
this fact is not in contradiction with our result. The point is that at 
nonzero 
temperature the symmetry between a spatial and a temporal loop is broken. 
't Hoofts argument \cite{thooft} 
then only rules out the simultaneous area law behaviour for 
the spatial 't Hooft loop and a 
timelike Wilson loop. In the plasma 
phase in fact the string tension for timelike
Wilson loops disappears, and our result is therefore only natural.

Strictly speaking our proof applies at very high temperatures where the 
perturbative
calculation of $<V(C)>$ is valid. However keeping in mind that at zero temperature
$V$ has a perimeter law behaviour the most natural possibility is that the 
change in the behaviour occurs at the deconfining phase transition. 
In fact the same argument by 't Hooft already mentioned before also
claims that either the
't Hooft loop or 
its dual must have 
an area law if no massless particles are present in the theory. Since one 
does not expect massless excitations 
to exist at any temperature, this 
line of reasoning would tell us that the dual 
string tension has to appear exactly 
at the same temperature at which the timelike Wilson loop ceases to
confine, i.e. at the phase transition.  

To
strengthen this argument we would like to present a simple 
discussion of why the area law of the 't Hooft loop is intimately related 
to the plasma like distribution of charges in the equilibrium state.

Consider an equilibrium neutral plasma of electric charges 
with the statistical sum
\begin{equation}
Z=\sum_{\{\rho\}}e^{-\beta\sum_{x,y}e^2\rho(x)D(x-y)\rho(y)}
\end{equation}
Here $\rho$ is the integer valued charge density 
\begin{equation}
\rho(x)=\sum_{x_i}n_i\delta(x-x_i)
\label{rho}
\end{equation}
The integer $n_i$ count how many positive or negative charges reside at the point $x_i$.
The interaction between the charges is via the Coulomb potential $D(x-y)$ 
\begin{equation}
D(x-y)=-<x|{1\over \partial^2}|y>
\end{equation}
The physics of this system is well 
known. Dynamically the screening length is generated and the ``dressed'' potential 
between the charges is screened on the distance scale $\xi$. For dilute plasma the 
screening length is inversely proportional to the fugacity $\mu$.
The convenient standard way to study this system is by duality transformation
\cite{samuel}:
\begin{equation}
Z=\int D\phi\sum_{\{\rho\}}e^{-{T\over 4e^2}\partial\phi\partial\phi +i\rho\phi} 
\label{dual}
\end{equation}
The scalar field $\phi$ ``classically'' is related to the charge density by
\begin{equation}
\partial_i^2\phi=-i2e^2\beta\rho
\end{equation}
Summing over $\rho$
this partition function is rewritten as
\begin{equation}
Z=\int D\phi e^{-{T\over 4e^2}\int 
(\partial_i\phi)^2+V(\phi)}
\label{pl}
\end{equation}
In the dilute plasma limit the potential 
\begin{equation}
V=\mu^2(1-\cos\phi)
\end{equation}
When the plasma is not dilute the explicit 
form of $V$ is not easy to calculate but
it is still true that $V$ 
is necessarily a periodic function of $\phi$ with period $2\pi$.
This periodicity is of course a direct consequence of the the integer valuedness
of the charge density eq.(\ref{rho}) and the way the field $\phi$ was introduced in 
eq.(\ref{dual})
Therefore $V$ 
has discrete minima and the classical equations for $\phi$  have wall like solutions.
Now let us calculate the average of the 'tHooft loop. As we have discussed earlier, 
the 't Hooft loop is a simple operator which measures the electric flux through
the area bounded by a contour $C$. The electric field is given in terms of 
the charge density as
\begin{equation}
\partial_iE_i=e\rho
\end{equation}
A short calculation shows that the electric flux through a contour $C$ due to a point
charge at the point $x$  
is proportional to the solid angle $\omega$.
Therefore with the appropriate normalization the 't Hooft loop operator is
\begin{equation}
V(C)=e^{{i\over 2}\int d^3x\rho(x)\omega_C(x)}
\end{equation}
It is now obvious that the calculation of the 't Hooft loop in this system is 
{\it verbatim} equivalent to Polyakov's calculation of the Wilson loop in the plasma of magnetic
monopoles \cite{polyakov}. The loop therefore has the area law with the dual string tension
equal to the wall tension of the classical wall solution of eq.(\ref{pl}).

This simple discussion is a very close caricature to the QCD calculation presented
in this note. Recall, that the electric field in 
QCD in the Euclidean finite temperature calculation is related to the scalar potential
by
\begin{equation}
E^i\propto i\partial_iA_0
\end{equation}
This together with the Gauss' law
\begin{equation}
\partial_iE_i=\rho\equiv g A_i\times E_i
\end{equation}
shows that $A_0$ is the exact counterpart of $\phi$ if one thinks of the charges that constitute
the plasma as the colour charged gluons.
More precisely, the QCD analog of the charge in the simple plasma calculation
is the hypercharge $Y$.
Of course, the QCD plasma at high temperature is not dilute, and as a result the potential
$V$ is not quite the simple sine-Gordon potential as in \cite{polyakov}. It does however
preserve the basic feature of periodicity.
The basic physics does not depend on the exact shape of the potential and therefore
as far as the 't Hooft loop is concerned it is the same at high temperature QCD
and in the simple equilibrium plasma of charges.

What happens in the 
confining phase? Of course the physics there is nonperturbative and
therefore no quantitative statements are available. It is however easy to understand
the basic features. Since there are no free charges, the fugacity in the confining
phase vanishes. Consequently 
there is no potential that suppresses fluctuations of $A_0$ (sic. $\phi$) - no 
Debye
mass, and the dual string tension vanishes.

We note  peripherically to our main discussion, that the plasma 
picture gives a natural explanation of a somewhat unusual scaling of the 
domain wall tension with $N$. The area law of the 't Hooft loop is due
to the presence in the plasma of gluons with nonzero hypercharge $Y$. There
are $N-1$ gluons with $Y=N$ and $N-1$ gluons with $Y=-N$, while the rest of 
the $(N-1)^2$ are hypercharge neutral. A gluon with the hypercharge 
$\pm N$
which sits close to the minimal area spanned by the contour $C$ contributes
a factor $-1$ to $<V>$. This is due to the fact that only half of the electric flux
emanating from this gluon, crosses the area of the loop. For $k$ gluons the
factor is obviously $(-1)^k$. 
Due to the plasma screening effect, only those gluons that are at a distance smaller 
than the screening length $1/m$ are effective in the disordering of the dual loop.
Let us for simplicity assume that the 
distribution of 
the number of gluons in the plasma 
 at distances scales smaller than the correlation length
$1/m$ is random and follows the Poisson distribution.
\begin{equation}
P(k)={\bar k^k\over k!}e^{-\bar k}
\end{equation}
Here $\bar k$ is the average number of gluons in the disk of thickness $1/m$
around the minimal area $S$ spanned by the loop.
The average of the loop is then estimated as 
\begin{equation}
<V>=\sum_kP(k)(-1)^k=e^{-2\bar k}
\end{equation}
Now
\begin{equation}
\bar k={1\over m}S_mn(T)
\end{equation}
where $n(T)$ is the density of the charged gluons in the plasma.
Since there are $2(N-1)$ species of charged gluons
the density in the large $N$ limit scales as 
\begin{equation}
n\propto N
\end{equation}
The screening length $1/m$ at large $N$ is finite, $m^2\propto g^2N$. 
The dual string/wall tension
therefore scales as
\begin{equation}
\alpha={1\over m}n\propto{N\over\sqrt{g^2N}}
\end{equation}
This is precisely the scaling found in the semiclassical 
calculation \cite{wall}.

The last point we want to discuss is the fate of the 't Hooft loop in a theory
which contains fundamental charges.
Here we expect $V$ to have an area law already at zero temperature which
presumably
then survives at arbitrary temperature. 
In fact the situation is slightly more intricate. Recall that our definition of $V$ eq.(\ref{v1})
did not
depend on the surface $S$ only in the absence of fundamental charges. In the presence of such 
charges, the operator $V$ depends explicitly not only on the contour $C$ but also 
on the surface
$S$ on 
which the angular function $\omega$ is defined to have the discontinuity. 
In fact it creates a current across the surface $S$ \cite{v}.
As a result one 
expects the average of $V$ will have an area law of the form
\begin{equation}
V_S(C)=\exp\{-\tilde\alpha S\}
\label{areal}
\end{equation}
where $S$ is not the minimal area subtended by $C$, but rather the area of the surface which
enters explicitly in the definition of $V$. The dual string 
tension $\tilde\alpha$ is due to the vacuum
fluctuations of the fundamental charges and should therefore be a decreasing 
function of the mass of the lightest fundamentally charged field.

At high temperature this expectation is easily confirmed by a simple analysis.
Consider as an example the theory with $n_f$ flavours of fundamental fermions.
Repeating our calculation we obtain the analog of eq.(\ref{action}) but with potential 
which is not periodic in $q$ and has therefore only one minimum at $q=0$\cite{korthals}. 
\begin{eqnarray}
V_{eff}&=&{4
\over 3}\pi^2T^4 
N^2(N-1)\left({q\over 2\pi}\right)^2\left(1-{Nq\over 2\pi}\right)^2\\
&-&{4
\over 3}\pi^2T^4 n_f\left[(N-1)\left({q\over 2\pi}+{1\over 2}\right)^2\left({q\over 2\pi}-{1\over 2}\right)^2
+\left({(N-1)q\over 2\pi}+{1\over 2}\right)^2\left({(N-1)q\over 2\pi}-{1\over 2}\right)^2\right]\nonumber
\end{eqnarray}
The minimum
at $q=\pi$ has now higher energy and therefore is only metastable.
Such a potential of course does not have a stable wall solution. Let us now consider the 
't Hooft loop operator which is defined on a surface  $z=f(x,y)$.
For simplicity let us assume that for $x$ and $y$ inside the contour $C$, the function $f$ 
is large and negative. 
Then the solution of the equations of motion we are looking for will be
very close to the vacuum 
value $q=0$ just to the left of the surface $S$. However, since the
solution is forced to 
have a discontinuity across $S$, the field will be equal to $\pi$ 
just to the right of $S$. 
Since the potential is not degenerate at $0$ and $\pi$, the region 
adjacent to the wall form the right has nonvanishing potential energy. Clearly in the
minimal action solution the field 
will tend to the vacuum value $q=0$ within a layer of
thickness of order of the Debye mass ``glued'' to the surface $S$. The action of such
a configuration is proportional $S_{eff}\propto \tilde\alpha S$. 

At low temperatures the semiclassical analysis is not adequate. However 
the fluctuations of the fundamental charges are still present in the vacuum. Since the
't Hooft loop creates the fundamental current through the surface $S$ it is clear 
that the vacuum in the presence of the loop is modified everywhere along $S$, and 
therefore the 't Hooft loop must have the same type of area law behaviour as 
in eq.(\ref{areal}).
We therefore conclude that, just like the 
Polyakov line, the 't Hooft loop ceases to be an order parameter in the strict sense: 
it has the 
area law behaviour on both sides of the deconfining phase transition. We expect
however, that again just
like with the Polyakov line, even though in the dual string tension $\alpha$ is nonvanishing 
everywhere,
it jumps strongly across the phase transition and therefore in practical sense should be a 
good indicator  of the transition. The intuitive basis for this expectation in QCD is the
chiral symmetry restoration in the plasma phase. Below the phase transition the quarks
have a dynamically generated mass, which disappears above the critical temperature. The dual 
string tension $\tilde\alpha$ must have inverse dependence on the quark mass, since it 
 vanishes in the confining phase for infinitely heavy quarks. 
Therefore one can expect a jump in $\tilde\alpha$
at the phase transition which is of order $\tilde\alpha$ itself.

Finally we note that the loop average can be measured 
on the lattice. 
The lattice version of $V$ is~\cite{groeneveld}:
\begin{equation}
V=\exp{\beta_L (1-\exp{i2\pi/N})\sum_{x\in S}\left({1\over N}TrP_{zt}(x)+c.c\right)}
\end{equation}  
where $\beta_L$ the lattice coupling
and  $P_{zt}$ are the electric plaquettes
orthogonal to the
surface.
It was precisely this operator, with the surface S covering {\it{all}} of a
given x-y cross section in a periodic volume, that was used by Kajantie
et al.\cite{wall}. Thus the dual string tension of the 
't Hooft loop and the wall tension measured by this group must be identical. 
It would be desirable to have results closer to the continuum limit.

{\bf Acknowledgements}
The work of A.K. is supported by PPARC, CNRS and a joint CNRS-Royal Society
project. 

\pagebreak


\begin{thebibliography}{99}


\bibitem{wall} T. Bhattacharya, A. Goksch, C.P. Korthals - Altes and R. Pisarski,
{\it Phys. Rev. Lett.}{\bf 66}(1991) 998, {\it Nucl. Phys.} {\bf B383} (1992) 497;   

\bibitem{larry} L. McLerran and B.  Svetitsky {\it Phys. rev.} {\bf D24} (1981) 450;

\bibitem{ben} B. Svetitsky and L. Yaffe {\it Phys. Rev.} {\bf D26} (1982) 693;
{\it Nucl. Phys.} {\bf B210} (1982) 423

\bibitem{perturb} 
P. Arnold , C. Zhai, {\it Phys.Rev.} {\bf D50} 7603 (1994);
{\it Phys.Rev.} {\bf D51} 1906 (1995); 
B. A. Freedman and L. D. McLerran, {\it Phys. Rev.} {\bf D16}, 1130 (1977); 
{\it ibid.}, {\bf D16}, 1147 (1977); 
{\it ibid.}, {\bf D16}, 1169 (1977); 
J. I. Kapusta, {\it Nucl. Phys.} {\bf B 148}, 461 (1979)

 

\bibitem{lattice} For a review see e.g. LATTICE 98, Proceedings of the 1998
Meeting on Lattice Gauge Theory, Ed T. Legrand, North-Holland Elsevier,
1999 . For new results: ibid., P. Vranas et al.,Nucl.Phys.Proc.Suppl.73:456-458,1999.

\bibitem{wallsfor} C.P. Korthals Altes, K. Lee and R.D. Pisarski, {\it Phys. Rev. Lett.} 
{\bf 73} (1994) 1754;
C.P. Korthals Altes and N.J. Watson, 
{\it Phys. Rev. Lett.} {\bf 75} (1995) 2799;
I. I. Kogan, {\it Phys. Rev.} {\bf D49}, 6799 (1994)

\bibitem{wallsagainst}V.M. Belyaev, I.I Kogan, G.W. Semenoff and N. Weiss
{\it Phys. Lett.} {\bf B277} (1992) 331; 
A. Smilga {\it Ann. of Phys.} {\bf 234} 
(1994), 1;.T. H. Hansson, H. B. Nielsen, and I. Zahed{\it Nucl. Phys.} {\bf B451}, 162 (1995)
W. Chen, M. I. Dobroliubov and G. W. Semenoff, {\it Phys. Rev.} {\bf D 46} 
(1992), 1223

\bibitem{lwall}K. Kajantie, L. K\"{a}rkk\"{a}inen, 
{\it Phys. Lett} {\bf B214}, 595 (1988);
K. Kajantie, L.  K\"{a}rkk\"{a}inen, and K. Rummukainen, 
{\it Nucl. Phys.} {\bf B333}, 100 (1990); {\it ibid.}, {\bf B357}, 693 (1991);
{\it Phys. Lett.} {\bf B286}, 125 (1992);
S. Huang, J. Potvin, C. Rebbi and S. Sanielevici, 
{\it Phys. Rev.} {\bf D42}, 2864 (1990); (E) 
{\it ibid.} {\bf D43}, 2056 (1991);
R. Brower, S. Huang, J. Potvin, C. Rebbi,
{\it ibid.} {\bf D46}, 2703 (1992),
R. Brower, S. Huang, J. Potvin, C. Rebbi, and J. Ross,
{\it ibid.} {\bf D46}, 4736 (1992);
W. Janke, B. A. Berg, M. Katoot, {\it Nucl. Phys.} {\bf B382}, 649 (1992);
B. Grossmann, M. L. Laursen, T. Trappenberg, and U. J. Wiese,
{\it ibid.} {\bf B396}, 584 (1993);
B. Grossmann and M. L. Laursen, {\it ibid.} {\bf B408}, 637 (1993);
Y. Aoki and K. Kanaya, {\it Phys. Rev.} {\bf D50}, 6921 (1994);
Y. Iwasaki, K. Kanaya, L. Karkkainen, K. Rummukainen, and T. Yoshie,
{\it ibid.} {\bf D49}, 3540 (1994);
S. T. West and J.F. Wheater, {\it Phys. Lett.} {\bf B383}, 205 (1996);
{\it Nucl. Phys.} {\bf B486}, 261 (1997);
K. Kajantie, M. Laine, A. Rajantie, K. Rummukainen, and M. Tsypin,
hep-lat/9811004.

\bibitem{mike}C. P. Korthals Altes, A. Michels, M. Stephanov, and M. Teper,
{\it Phys. Rev.} {\bf D55} 1047 (1997);

\bibitem{thooft}G.'t Hooft, {\it Nucl. Phys.} {\bf B138}, 1 (1978)

\bibitem{weiss} N. Weiss {\it Phys. Rev.}{\bf D24} (1981) 475; {\bf D25} (1982) 2667;D. Gross, R. D. Pisarski, and L. G. Yaffe, {\it Rev. Mod. Phys.}{\bf 53}, 497 (1981). 

\bibitem{korthals} C.P.Korthals Altes, {\it Nucl. Phys.} {\bf B420} (1994) 637;

\bibitem{samuel}  S. Samuel, {\it Phys.Rev.} {\bf D18} (1978),1916;

\bibitem{polyakov} A.M. Polyakov, {\it Nucl.Phys.} {\bf B120}, 429 (1977) 

\bibitem{v} A. Kovner and B. Rosenstein, {\it Int. J. Mod. Phys.} {\bf A7} (1992) 7419; 

\bibitem{groeneveld}J. Groeneveld, J. Jurkiewicz and C.P. Korthals Altes,
 {\it Physica Scripta} {\bf 23}, Nr. 5 : 2, p 9  1022 (1981)

\end{thebibliography}
\end{document}